\documentclass[12pt,eqsecnum,preprint]{aastex}
\title{Inertial Current Generators of Poynting Flux in MHD Simulations of Black Hole Ergospheres}
\begin{document}
\author{Brian Punsly}
\affil{4014 Emerald Street No.116, Torrance CA, USA 90503 and
International Center for Relativistic Astrophysics,
I.C.R.A.,University of Rome La Sapienza, I-00185 Roma, Italy}
\email{brian.m.punsly@boeing.com or brian.punsly@gte.net}
\begin{abstract}This Letter investigates the physics that is responsible for
creating the current system that supports the outgoing Poynting
flux emanating from the ergosphere of a rotating black hole in the
limit that the magnetic energy density greatly exceeds the plasma
rest mass density (magnetically dominated limit). The underlying
physics is derived from published three-dimensional simulations
that obey the general relativistic equations of perfect
magnetohydrodynamics (MHD). It is found that the majority of the
Poynting flux emitted from the magnetically dominated regions of
the ergosphere has a source associated with inertial effects
outside of the event horizon.
\end{abstract}
\section{Introduction}There are two known theoretical mechanism for producing
field-aligned outgoing poloidal Poynting flux, $S^{P}$, at the
expense of the rotational energy of a black hole in a
magnetosphere that is magnetically dominated. There are
electrodynamic processes collectively called Blandford-Znajek
mechanisms in which currents flow virtually parallel to the proper
magnetic field direction (force-free currents) throughout the
magnetically dominated zone all the way to the event horizon
\citep{blz77,phi83,thp86}. Therefore, these electrodynamic
currents have no source within the magnetically dominated black
hole magnetosphere. Alternatively, there is the GHM
(gravitohydromagnetic) dynamo in which large relativistic inertia
is imparted to the tenuous plasma by black hole gravity that in
turn creates a region of strong cross-field currents (inertial
currents), $J^{\perp}$, that provide the source of the
field-aligned poloidal currents, $J^{P}$, that support $S^{P}$ in
an essentially force-free outgoing wind, i.e.,
$\nabla\cdot\mathbf{J}\approx\partial J^{\perp}/\partial X^{\perp}
+ J^{P}/\partial X^{P}\approx 0$ at the source \citep{pun01}. The
force-free electrodynamic current flow is defined in terms of the
Faraday tensor and the current density as $F^{\mu\nu}J_{\nu}=0$.
As a consequence, $\mathbf{J}\cdot\mathbf{E}=0$, so $S^{P}$ in a
force-free magnetosphere must be injected from a boundary surface.
The two types of sources associated with these two types of
Poynting fluxes are quite distinct in the ergopshere: the inertial

current provides the $\mathbf{J}\cdot\mathbf{E}$ source in
Poynting's Theorem and the force-free (electrodynamic) component
of $S^{P}$ emerges from a boundary source at the event horizon.
\par Numerical models can be useful tools for understanding the source of $S^{P}$
emerging from the ergosphere of a black hole magnetosphere. Some
recent three-dimensional simulations in
\citet{dev03,hir04,dev05,dev06,kro05} show $S^{P}$ emanating from
magnetically dominated funnels inside of the vortices of thick
accretion flows. In these simulations, $J^{\perp}\approx
J^{\theta}$ (in Boyer-Lindquist coordinates which are used
throughout the following) inside of the ergosphere, since the
poloidal field settles to a nearly radial configuration early on
in the simulation \citep{hir04}. In principle, one can clearly
distinguish the amount of $S^{P}$ emerging form the ergosphere in
a simulation that is of electrodynamic origin (as proposed in
\citet{blz77}) from the amount due inertial effects (the GHM
theory of {\citet{pun01}) by quantifying the relative strengths of
$S^{P}$ emerging from the inner boundary (the asymptotic
space-time near the event horizon) with the amount created by
sources within the ergosphere. In the high spin rate simulations
in questions, over 70\% of $S^{P}$ emerging from the ergopsheric
funnel is created outside of the inner boundary.
\section{Physical Quantities in Boyer-Lindquist Coordinates}
The Kerr metric (that of a rotating uncharged black hole),
$g_{\mu\nu}$, in Boyer-Lindquist coordinates $(r,\theta,\phi,t)$,
is given by a line element that is parameterized by the black hole
mass, $M$, and the angular momentum per unit mass, $a$, in
geometrized units \citep{thp86}. We use the standard definitions,
$\rho^{2}=r^{2}+a^{2}\cos^{2}\theta$ and $\Delta =
r^{2}-2Mr+a^{2}$, where $\Delta=0$ at the event horizon,
$r_{_{+}}=M+\sqrt{M^{2}-a^{2}}$. The "active" region of space-time
is the ergosphere, where black hole energy can be extracted,
$r_{_{+}}<r <r_s = M + \sqrt{M^2 - a^2 \cos^2 \theta}$
\citep{pen69}.
\subsection{The Toroidal Magnetic Field Density and the Cross-Field EMF}
The flux of electromagnetic angular momentum along the poloidal
magnetic field direction is the component of the stress-energy
tensor, $T_{\phi}^{\; r}= [1/(4\pi)]F_{\phi\alpha}F^{\alpha r}$,
in the approximation that the field is radial. In steady state,
the electromagnetic angular momentum flux per unit poloidal
magnetic flux is the toroidal magnetic field density:
$-B^{T}\equiv\sqrt{-g}\,F^{\theta r}$, where
$g=-\rho^{4}\sin^{2}{\theta}$ \citep{phi83,pun01}. Similarly, the
electromagnetic energy flux along the poloidal magnetic field
direction, $S^{P}$, is the component, $T_{t}^{\; r}=
[1/(4\pi)]F_{t \alpha}F^{\alpha r}$, in the approximation that the
poloidal field is radial \citep{thp86}. In steady state, the
energy flux per unit poloidal magnetic flux is
$-(\Omega_{_{F}}/c)B^{T}$, where $\Omega_{_{F}}$ is the field line
angular velocity \citep{phi83,pun01}. Consequently, $B^{T}$ is
useful for quantifying the energy and angular momentum fluxes as
steady state is approached. From Ampere's law,
\begin{eqnarray}
\sqrt{-g}J^{\theta}=B^{T}_{\;\; ,r}+(\sqrt{-g}F^{\theta
t})_{,t}\;.
\end{eqnarray}
At late times, as an approximate steady state is reached, one
expects $\sqrt{-g}J^{\theta}\approx B^{T}_{\;\; ,r}$. Therefore,
at late times, $J^{\theta}$ is a potential source for the current
system that supports $S^{P}$.
\par Even when a system
has not reached a time stationary state, one can introduce a
well-defined notion of $\Omega_{_{F}}$ that becomes the field line
angular velocity in the steady state. If the field is nearly
radial one can simply define the expression, $F_{t\theta}\equiv
 -\Omega_{_{F}}F_{\theta\phi}$. With this definition, $\Omega_{_{F}}$
is a function of space and time and in steady state it becomes a
constant along a perfect MHD flux tube. Therefore, the EMF across
the magnetic field is $-\Omega_{_{F}}F_{\theta\phi}$ by
definition.
\subsection{The Source of Poloidal Poynting Flux}In
Boyer-Lindquist coordinates, the curved space-time equivalent of
the "$\mathbf{J}\cdot\mathbf{E}$" source of $S^{P}$ is the term
$F_{t\alpha}J^{\alpha}$. This notion is described by the integral
version of Poynting's Theorem which is the integral of
$T^{\nu}_{t\;\; ;\nu}=F_{t\nu}J^{\nu}$ combined with Stokes'
theorem \citep{thp86,phi83}
\begin{eqnarray}
&&\int [F_{t\phi}J^{\phi}+F_{tr}J^{r}+F_{t\theta}J^{\theta}]\, dV
- \frac{d}{dt}\int T^{\,t}_{t}\, dV =[1/(4\pi)]\oint \sqrt{-g}F_{t
\alpha}F^{\alpha n}\;dA;,
\end{eqnarray}
where $dV=\sqrt{-g}dr d\phi d\theta$ is the spatial volume element
of a section of a thick spherical shell and $-T^{\,t}_{t}$ is the
energy density of the field and $n$ is the normal direction to the
surface area element, $dA$, of the Gaussian pillbox. In steady
state, the source of $S^{P}$ is
$-\Omega_{_{F}}F_{\theta\phi}J^{\theta}$ in the approximation of a
radial field. Without the radial approximation, one needs to
define a poloidal field direction, $\mathbf{B^{P}}$, and a
cross-field poloidal EMF, $E_{\perp}\equiv
 -\Omega_{_{F}}B^{P}$ then the source of $S^{P}$ is
$E_{\perp}J^{\perp}$. This can all be setup with great
mathematically complexity (see \citet{pun01} for sample
calculations) and no additional physical insight. The reader
should remember that the poloidal field is not exactly radial in
the following and the simple notion of $J^{\theta}$ is used to
approximate $J^{\perp}$.
\section{The Source of Ergospheric Poynting Flux in the KDE Simulation}
The simulation "KDE" of \citet{dev03,hir04,dev05,dev06,kro05} is
of the most interest since it generates an order of magnitude more
$S^{P}$ than any of the other simulations \citep{kro05}. The
magnetically dominated funnel spans the latitudes $0^{\circ} <
\theta < 55^{\circ}$ at the inner calculational boundary, near
$r_{_{+}}$. Quantifying the magnetic dominance in the funnel is
the pure Alfven speed, $U_{A}=B^{P}/(\sqrt{4\pi n \mu}c)$, where
$n$ is the proper number density, $\mu$ is the specific enthalpy
of the plasma and $B^{P}$ is the poloidal field strength. Within
the funnel $10 < U_{A}^{2} < 10^{4}$. The funnel $S^{P}$ averaged
over time and azimuth in KDE near $r_{_{+}}$ is shown in figure 1.

This contour map indicates a region of strong outgoing $S^{P}$
 in the evacuated funnel, $30^{\circ} < \theta < 55^{\circ}$,
$r\gtrsim r_{_{+}}$. It is clear that $S^{P}$ suddenly diminishes
close to $r_{_{+}}$ at $r\approx 1.3M-1.5M$. Inspection of the
contour map indicates that over $72\%$ of $S^{P}$ is created
within a thin layer near $r\approx 1.4 M$. Because of the
saturation of the dark red color in the plotting routine, $S^{P}$
might be even larger above the switch-off layer than indicated in
the contour map. Thus, we only have a lower bound on the strength
of $S^{P}$ above the switch-off layer and it is likely that more
than 72\% of the energy flux is created within this thin layer.
This effect even continues into the weak $S^{P}$ region closer to
the pole at $20^{\circ} < \theta < 30^{\circ},\: r\gtrsim
r_{_{+}}$.
\par In order to investigate possible source terms for $S^{P}$,
in (2.2), one needs to be explicit about what is plotted in figure
1. Whenever a time lapse of 80 M occurs within the high time
resolution simulation, data is stored for a time snapshot. There
are 75 of these snapshots that are averaged in figure 1, steps 26
through 100. Thus, the discrete time average of (2.2) is relevant
,
\begin{eqnarray}
&&\frac{1}{75}\sum_{i=26}^{100}\left[\int F_{t\alpha}J^{\alpha}\,
dV \right]_{i}-
\frac{1}{75}\sum_{i=26}^{100}\left[\frac{d}{dt}\int T^{\,t}_{t}\,
dV\right]_{i}
\nonumber\\
&& =[1/(4\pi)]\frac{1}{75}\sum_{i=26}^{100}\left[\int_{1+3}
\sqrt{-g}F_{t \alpha}F^{\alpha r}\,d\theta\,d\phi+\int_{2+4}
\sqrt{-g}F_{t \alpha}F^{\alpha \theta}\,dr\,d\phi\right]_{i}\;,
\end{eqnarray}
where the four sides of the Gaussian pillbox of integration in
figure 1 are the four curves labelled "1 -4". The result of
significance from this plot is that $\sum_{i=26}^{100}\int_{3}
\sqrt{-g}F_{t \alpha}F^{\alpha r}\,d\theta\,d\phi >
2\mid\sum_{i=26}^{100}\int_{1} \sqrt{-g}F_{t \alpha}F^{\alpha
r}\,d\theta\,d\phi\mid$, i.e. the time averaged
 $S^{P}$ increases across this thin volume. There are three
 possible source terms for this increase in $S^{P}$ when (3.1) is applied to the pillbox in
 figure 1. The time averaged field decay from a finite spatial region
 (the second term on the LHS) is not a physically viable
 source of long term $S^{P}$ (unless the simulation is pathological and keeps creating
 local field energy from numerical artifacts).
 The only reasonable
 choices are the $\mathbf{J}\cdot\mathbf{E}$ term and the
 conversion of latitudinal Poynting flux, $S^{\perp}$, from the surface terms
 $\sum_{i=26}^{100}\int_{2+4}\sqrt{-g}F_{t \alpha}F^{\alpha \theta}\,dr\,d\phi$. We don't have
time averaged plots of $S^{\perp}$, so we don't know how the
magnitude of this term. Because of the symmetry at the pole, the
only viable scenario is that $S^{\perp}$ radiates poleward from
its source in the funnel wall and somehow converts to $S^{P}$ in
the pillbox. One thing that is unequivocal from figure 1, is that
during the lifetime of the simulation, less than $30\%$ of $S^{P}$
that reaches emerges from the ergospheric funnel came from the
inner boundary, the rest was created external to the boundary.
\par The existence of this putative source
is a highly significant result and it is desirable to check it by
other means. For example, since this occurs near the boundary how
do we know that this is not an artifact of an inherent error in
the plotting routine? Figure 7 of \citet{kro05} is a similarly
averaged (in time and azimuth) plot of the magnetic angular
momentum flux, $T_{\phi}^{\; r}= [1/(4\pi)]F_{\phi\alpha}F^{\alpha
r}$. Figure 7 shows that $T_{\phi}^{\; r}$ is created
predominantly in the ergosphere, over 70\% of the electromagnetic
angular momentum flux in the funnel is created by sources within
the ergopshere for $25^{\circ} < \theta < 55^{\circ}$. The
remaining fraction can be associated with electrodynamic sources
(i.e., sourceless and emerging from the inner boundary). A
consistent picture emerges from the time average of Ampere's law
in (2.1) if the dominant source term for $S^{P}$ in (2.2) is
$F_{t\alpha}J^{\alpha}\approx F_{t\theta}J^{\theta}$.
\section{The Cross-field Current Density in the Ergosphere} In this section, we use a plot of
the strong electromagnetic forces within the evacuated funnel to
try to understand the physical mechanism that drives the source of
$S^{P}$ in the simulation, KDE. Presently, there is no existing
data that has been extracted from the KDE simulation that directly
illustrates the electromagnetic forces or currents in the
ergosphere. Fortunately, plots of the electromagnetic force
already exist for the KDP model of \citet{dev05} which is
characterized by $a/M =0.9$. So, the best we can do is to explore
electromagnetic force plots (such as figure 2) from a closely
related simulation, KDP. This model has the second highest $S^{P}$
luminosity within the family of simulations, but it is an order of
magnitude weaker than the KDE model. This should still be
qualitative adequate for the following analysis, since it is
stated in \citet{dev05} that the azimuthal force exists in all the
models in the same relative location, but it is strength
correlates with spin and is the most pronounced in the KDE model.
\par The evacuated funnel of KDP is dominated by
magnetic energy, $10^{4}>U_{A}^{2}>10$ (according to figure 3 of
\citet{hir04}) in the ergosphere if $0^{\circ} < \theta <
65^{\circ}$. The nonforce-free nature of the current density will
be studied by means of a plot of the electromagnetic force (that
were presented in \citet{dev05} and does not appear in
\citet{dev06}), $F^{\phi\nu}J_{\nu}=F^{\phi t}J_{t}+ F^{\phi
r}J_{r}+F^{\phi\theta}J_{\theta}$, within the funnel at $t=8080M$,
averaged over azimuthal angle.

One can scale the individual components of the strong azimuthal
force that appears at $r\approx 1.5M$ in figure 2 if turbulence
does not dominate the dynamics: $F^{\phi t}J_{t}\sim
\alpha^{-2}F_{\phi t}$, $F^{\phi r}J_{r}\sim\alpha^{-2}F_{r t}$
and $F^{\phi\theta}J_{\theta}\sim\alpha^{-2}F_{\theta t}$, where
the quantity, $\alpha
=\sqrt{\Delta}\sin{\theta}/\sqrt{g_{\phi\phi}}$, is the lapse
function that represents a global redshift factor
\citep{thp86,pun01}. The lapse function vanishes at the horizon
and is therefore useful in expansions in a small dimensionless
parameter near the horizon (at r=1.5M, $\alpha^{-2}=58.8$,
$\theta=45^{\circ}$). These scalings yield the approximation
\begin{eqnarray}
&& F^{\phi\nu}J_{\nu}\approx
F^{\phi\theta}J_{\theta}\approx\frac{2Mra}{c\rho^{4}\Delta}\left(\Omega_{_{F}}-\Omega\right)F_{\phi\theta}J^{\theta}\;,
\end{eqnarray}
where $\Omega=-g_{\phi t}/g_{\phi\phi}$ is known as the ZAMO
angular velocity and it approaches the horizon angular velocity
deep in the ergosphere (see \citet{thp86} for more details).
Figure 2 indicates large changes in $F^{\phi\nu}J_{\nu}$ occur on
the order of the grid size at $r\approx 1.5M$, $\sim 0.01M$. Thus,
this must be a dynamic effect and does not derive from
$\nabla\alpha^{-2}$. A consistent picture can be constructed by a
large $J^{\theta}$, enhanced by a factor of more than 20 compared
to the upstream flow, that is responsible for the large force that
initiates at $\alpha\approx 0.10-0.25$. From Ampere's law in
(2.1), this should be a prestigious source of $B^{T}$ and by (2.2)
this should also be the strong source of $S^{P}$,
$F_{t\theta}J^{\theta}$ consistent with the analysis of the KDE
simulation in section 2. Note that
$F_{\phi\alpha}J^{\alpha}\approx F_{\phi\theta}J^{\theta}$ is the
force that torques the plasma \citep{thp86,pun01}. So, the
Poynting flux generation in this scenario is associated with a
strong electromagnetic torque.
\par A physical explanation of the strong electromagnetic torque is the
most basic concept of GHM.

Namely, the field is rotating a rate that is slower than the
enforced rotation velocity of the plasma that is dictated by the
dragging of inertial frames in the ergosphere. Gravity tries to
pull the plasma forward relative to the field as the horizon is
approached, the back reaction of the field is to torque the plasma
backwards back onto Larmor helices that thread the field lines. In
the process, the field is partially overwhelmed by gravity and
twisted forward azimuthally creating $B^{T}<0$ even though the
plasma is tenuous. Note that $J^{\theta}$ directed equator-ward in
the northern hemisphere of the KDE simulation is of the correct
orientation to both source $B^{T}<0$ and to torque the plasma if
$B^{P}>0$ (for more details of this process see
\citet{pun01,sem04}). The strong force occurs in figure 2
throughout the range $10<U_{A}^{2}<10^{4}$ indicating that the
current is not driven across $B^{P}$ by the rest mass inertia, but
by a relativistic effect imposed on the tenuous plasma by the
black hole. Finally, GHM naturally explains the current closure
within the global wind. Figure 3 shows a remarkable agreement
between the current system of the analytical GMH model depicted in
figure 9.12 of \citet{pun01} and the KDE simulation. In the dynamo
source, $\partial(\sqrt{-g}J^{\theta})/\partial\theta +
\partial(\sqrt{-g}J^{r})/\partial r\approx 0$, where $J^{r}$ is the field
aligned current (dashed contours in figure 3) that support
$S^{P}$.
\section{Conclusion} In this paper, we studied simulations of a magnetically dominated
funnel of a rapidly rotating black hole, $a/M=0.998$. A source
that is responsible for creating over 70\% of $S^{P}$ transported
through funnel during the life of the simulation was found.
Similarly, one can conclude that $<30\%$ of $S^{P}$ emerging from
the ergospheric funnel is from an inner boundary source, near the
horizon. The small residual $S^{P}$ injected from the boundary
into the accretion wind can be considered of electrodynamic
origin. The distribution of $S^{P}$ in figure 1 is in contrast to
the Blandford-Znajek solution in which essentially all the $S^{P}$
is of electrodynamic origin, i.e., it emanates from the horizon
and passes through the accretion wind with minimal interaction,
thereby maintaining a virtually constant value along each poloidal
flux tube throughout the ergosphere.
\par There were two possible sources for $S^{P}$ in the funnel, there are the GHM inertial
currents and $S^{\perp}$ injected from the funnel wall. The latter
would be a new source of $S^{P}$ in a magnetically dominated
magnetosphere that has not been considered in previous literature.
The putative $S^{\perp}$ is created by inertial effects in the
funnel wall. We were unable to distinguish between these two
inertial sources with the available data. It was demonstrated that
a GHM current explains the ergospheric sources of $S^{P}$, the
electromagnetic angular momentum flux, the large azimuthal forces
seen in the ergosphere and global current closure in the wind
zone.
\par Further analysis of three-dimensional simulations about
rapidly rotating black holes ($a/M = 0.998$) are needed to clarify
the physics that creates $S^{P}$. At least three consecutive time
snapshots are needed in order to find the current distribution
from Ampere's Law at each coarse time step data dump. Since the
flow is highly turbulent near the event horizon, it would be
important to then time average the current distribution. Higher
resolution might be required to resolve the issue of whether the
source in figure 1 is from $J^{\perp}E_{\perp}$, $S^{\perp}$ or
some unexpected numerical error.
\section{Figure Captions}
Figure 1. The azimuthally averaged and time averaged (over 75\% of the simulation
    that ends at $t=8080 M$) Poynting flux from the
    model KDE ($a/M=0.998$) of \citet{kro05}. The figure is a magnification of the
    inner region of figure 10 of \citet{kro05}. It is an excision of a region,
    $0^{\circ} < \theta < 65^{\circ}, r\gtrsim r_{_{+}}$ that is a little
    larger than the ergospheric portion of the magnetically dominated
    funnel, $0^{\circ} < \theta <
55^{\circ}, r\gtrsim r_{_{+}}$. The majority of $S^{P}$
switches-off in a thin layer near $r=1.3M$
    - $r=1.5M$ (see \citet{kro05} for a description of the units on the color bar). This
    region is a source for the majority of the outgoing
    $S^{P}$ emerging from the funnel. A Gaussian pillbox is drawn as a dashed contour for
    use in Poynting's Theorem. There are 26 grid zones between the
    inner boundary, $r=1.175M$ and $r=1.5M$. The plot is provided courtesy of John Hawley
\par Figure 2. The azimuthally averaged azimuthal electromagnetic
    force in the evacuated funnel of the
    model KDP from the central region of figure 7 from \citet{dev05} at time $t=8080 M$ in geometrized units. Magnitude
    information is expressed by color in code units (see scale above). The funnel wall
    is marked by the solid white line. The azimuthal gas pressure forces are much smaller
    in this region \citep{dev05}. Note that there is a strong electromagnetic torque
    between the boundary at $r=1.45M$ and $r=1.75M$, an order of magnitude stronger
    than anywhere else in the ergosphere and the adjacent space-time.
    The plot only covers the funnel in the ergopshere in the
restricted span, $45^{\circ} < \theta < 65^{\circ}$, yet the full
range of Alfven speeds is captured, $10^{4}>U_{A}^{2}>10$.
    The plot is provided courtesy of John Hawley
\par Figure 3.  An overlay of the time stationary current system from a GHM
    model reproduced from figure 9.12 of \citet{pun01} and the Poynting flux distribution of KDE.
    The relative scales are set by the black hole (gray) radius, $r_{_{+}} =1.06M$, and the inner boundary of KDE at 1.175 M.
    The GHM model assumes very different initial conditions than KDE, but the basic current
    topology should be common to all GHM magnetospheres: an enhanced strong cross-field current
    density (which increases equator-ward) in the ergosphere that is the source of a field aligned current system in a
    nearly force-free wind zone and a strong return current flow at the edge of the wind zone

\end{document}